\documentclass[twocolumn,superscriptaddress]{revtex4}
\usepackage{graphicx}
\graphicspath{{Figs/}} %
\usepackage{color}
\usepackage{amssymb,amsfonts,amsmath}

\newcommand{\beginsupplement}{%
        \setcounter{table}{0}
        \renewcommand{\thetable}{S\arabic{table}}%
        \setcounter{figure}{0}
        \renewcommand{\thefigure}{S\arabic{figure}}%
     }

\newcommand{\beqn}{\begin{eqnarray}}
\newcommand{\eeqn}{\end{eqnarray}}
\newcommand{\beq}{\begin{equation}}
\newcommand{\eeq}{\end{equation}}

\renewcommand{\>}{\rangle}

\definecolor{junglegreen}{rgb}{0.16, 0.67, 0.53}
\definecolor{myrtle}{rgb}{0.13, 0.26, 0.12}
\definecolor{lincolngreen}{rgb}{0.11, 0.35, 0.02}
\definecolor{forestgreen}{rgb}{0.13, 0.55, 0.13}

\newcommand{\lastequal}{These authors contributed equally. Please send
  correspondance to \url{awalczak@lpt.ens.fr}, \url{tmora@lps.ens.fr}}

\newcommand{\LPENS}{Laboratoire de physique de l'\'Ecole normale sup\'erieure (PSL University), CNRS, Sorbonne Universit\'e, and Universit\'e de Paris, 75005 Paris, France}

\begin{document}
\title{Multi-lineage evolution in viral populations driven by host immune systems}

\author{Jacopo Marchi}
\affiliation{\LPENS}
\author{Michael L\"assig}
\affiliation{Institute of Theoretical Physics, University of Cologne, 50937 Cologne, Germany}
\author{Thierry Mora}
\affiliation{\LPENS}
\affiliation{\lastequal}
\author{Aleksandra M. Walczak}
\affiliation{\LPENS}
\affiliation{\lastequal}

\date{\today}

\begin{abstract}
 % abstract
Viruses evolve in the background of host immune systems that exert selective pressure and drive viral evolutionary trajectories. This interaction leads to different evolutionary patterns in antigenic space. Examples observed in nature include the effectively one-dimensional escape characteristic of influenza A and the prolonged coexistence of lineages in influenza B. Here we use an evolutionary model for viruses in the presence of immune host systems with finite memory to delineate parameter regimes of these patterns in a in two-dimensional antigenic space. We find that for small effective mutation rates and mutation jump ranges, a single lineage is the only stable solution. Large effective mutation rates combined with large mutational jumps in antigenic space lead to multiple stably co-existing lineages over prolonged evolutionary periods. These results combined with observations from data constrain the parameter regimes for the adaptation of viruses, including influenza.

\end{abstract}

\maketitle

\section{Introduction}

 % intro
Different viruses exhibit diverse modes of evolution~\cite{Gog2002, Gandon2016a, Yan2018, Koelle2009}, from relatively slowly evolving viruses that show stable strains over many host generations such as measles~\cite{Grenfell2002}, to co-existing serotypes or strains such as noroviruses~\cite{White2014} or influenza B~\cite{Rota1990,Bedford2012}, to quickly mutating linear strains such as most known variants of influenza A~\cite{Smith2004}. Despite the different patterns of evolutionary phylogenies and population diversity, all viruses share the common feature that they co-evolve with their hosts' immune systems. The effects of the co-evolution depend on the mutation timescales of the viruses and the immune systems, the ratio of which varies for different viruses. However, in the simplest setting, the population of hosts exerts a selective pressure on the viral population, {generating viral evolution towards increasing antigenic distance from the host population}. Here, we explore this mutual dynamics in a model of viruses that evolve in the background of host immune systems. While a lot of previous studies of pathogen-immune dynamics have foccussed on specific systems~\cite{Grenfell2002, Keeling2009, Bedford2012, Reich2013, Gandon2016a, BenShachar2014, Koelle2009, Boni2006, Luksza14, Fonville2014, OReilly2018}, we follow in the steps of more general considerations~\cite{Rouzine2018,Yan2018}. Specifically, we are interested in how the host immune cross-reactivity and memory control the patterns of viral diversity.

These evolutionary processes lead to a joint dynamics that has often been modeled by  so called Susceptible-Infected-Recovered (SIR) approaches to describe the host population~\cite{Kermack1927,AndersonMay1991}, possibly coupled with a mutating viral population. In their simplest form, these models have successfully explained and predicted the temporal and historical patterns of infections, such as measles~\cite{Grenfell2002}, where there are little mutations, or dengue, where enhancement between a small number of strains can lead to complex dynamics~\cite{Ferguson1999}. These methods have been important in helping develop vaccination and public health policies. 

Apart from a huge interest in the epidemiology of viruses~\cite{Keeling2009}, a large extension of SIR models has also tackled questions on the role of complete and partial cross-coverage, and how that explains infection patterns for different viruses~\cite{Reich2013, Gandon2016a}, the role of spatial structure on infections~\cite{Bedford2012}, as well as antigenic sin~\cite{BenShachar2014,Mongkolsapaya2003}. Most of these questions were asked with the goal of explaining infection and evolutionary patterns of specific viruses, such as dengue~\cite{Reich2013,BenShachar2014}, influenza~\cite{Koelle2009, Boni2006, Luksza14, Fonville2014} or Zika~\cite{OReilly2018}. Here we take a more abstract approach, aimed at understanding the role of cross-reactivity and mutation distance in controlling the evolutionary patterns of diversity.

At the same time, the wealth of samples collected over the years, aided by sequencing technologies, has allowed for data analysis of real evolutionary histories for many types of viruses. One of the emerging results is the relatively low dimensionality of antigenic space -- an effective phenotypic space that recapitulates the impact of hosts immune systems on viral evolution. Antigenic mapping, which provides a methodology for a dimensionality reduction of data~\cite{Smith2004} based on phenotypic titer experiments, such as Hemaglutanin Inhibition (HI) assays for influenza~\cite{Hirst1943}, has shown that antigenic space is often effectively low-dimensional.  For example, influenza A  evolution is centered on a relatively straight line in antigenic space~\cite{Fonville2014}. This form suggests that at a given time influenza A strains form a quasispecies of limited diversity in antigenic space, with  escape mutations driven by antigenic pressure moving its center of mass~\cite{Bedford2012,Rouzine2018,Yan2018}.

 {We focus on a simplified model of viral evolution in a finite-dimensional space that delineates evolutionary patterns with different complexity of coexisting lineages.  Recent models of these dynamics have focused only on the linear evolutionary regime relevant of influenza A~\cite{Bedford2012} or have used an infinite-dimensional representation of antigenic space~\cite{Yan2018}. Here we also model immune memory in more detail, while keeping a simplified infection dynamics with a small number of model parameters. While our treatment does not account for many features of host-immune dynamics (as discussed in sections~\ref{methods} and ~\ref{discussion}) it offers a stepping stone to future more in-depth analysis of the role of host repertoires. }

\section{Methods}~\label{methods}

 % methods
\subsection{The model}

\begin{figure*}
\centering
\includegraphics[width=.8\linewidth]{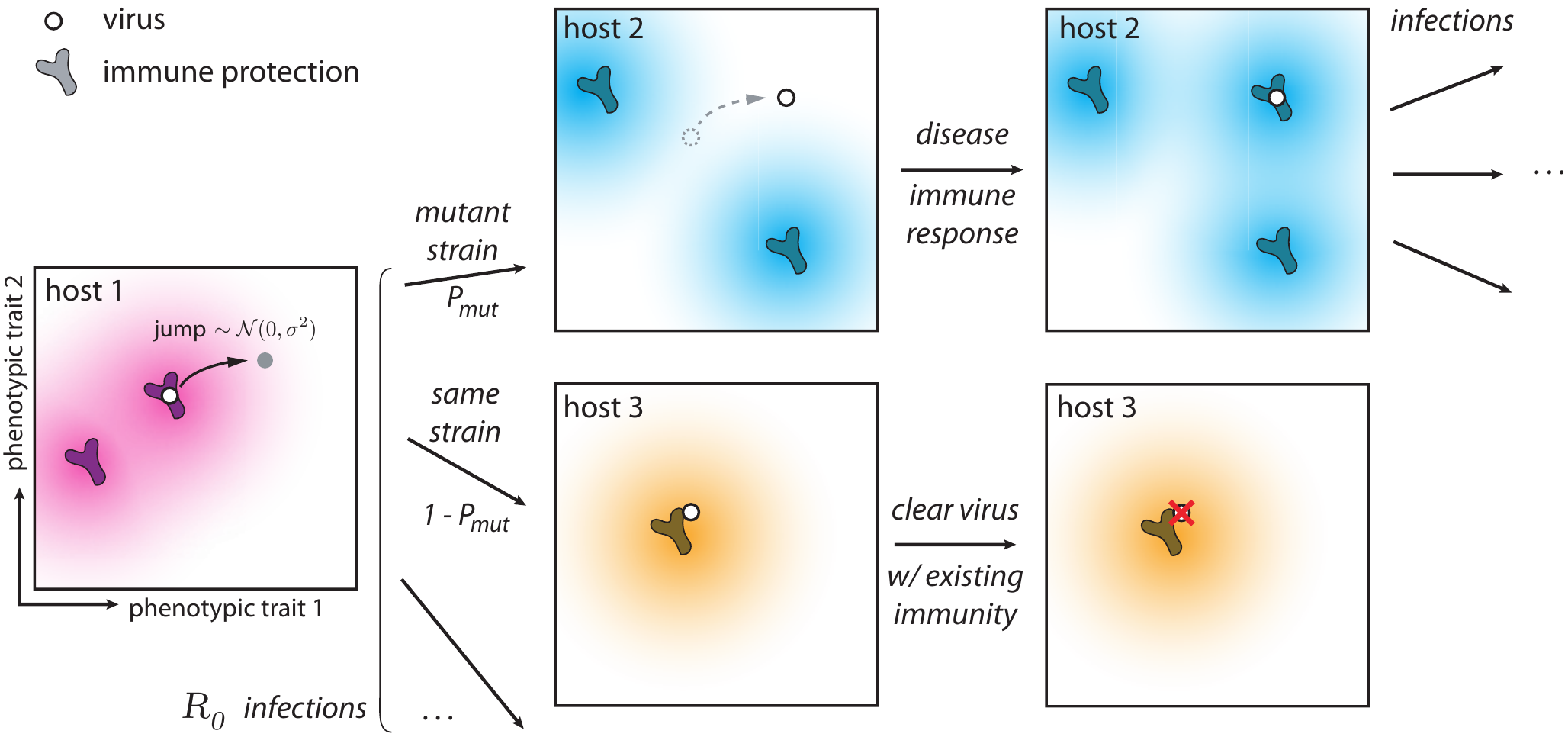}
\caption{\label{cartoon}{\bf Phenotypic space and key ingredients of the evolutionary model.} During an infection, a virus attempts to infect on average $R_0$ hosts, however not all infections are successful. The immune repertories of some hosts can clear the virus (case of host 3) since their cross-reactivity kernels from existing memory receptors confer protection. However if the host does not have protection against the infecting virus (case of host 2), the hosts becomes infected. After the infection this host acquires immunity against the infecting virus. Since the virus can mutate within a given host (host 1), the infecting virus can be a mutated variant (case of host 2) with probability $P_{\rm mut}=1-e^{-\mu t_I}$ and the ancestral strain that infected host 1 with rate $1-P_{\rm mut} =e^{-\mu t_I}$ (case of host 3). The cross-reactivity kernel is taken to be an exponential function $f(r) = {\rm exp}(- \frac{r}{d})$,
meaning that viruses are recognized by receptors if they are closer in phenotypic space.
Jumps are in a random direction and their size are distributed as a Gamma distribution of mean $\sigma$ and shape parameter 2.
The dimensionless raio $\sigma/d$ controls the ability of viruses to escape immunity.
We assume no selection  within one host.}
\end{figure*}

We implement a stochastic agent based simulation scheme to describe viral evolution in the background of host immune systems. Its main ingredients are sketched in Fig.~\ref{cartoon}. We fix the number of hosts $N=10^{7}$ and do not consider host birth-death dynamics. Host can get infected by a given viral strain if they are not already infected by it (equivalently to {\it susceptible} individuals in SIR models) in a way that the infection probability depends on the hosts infection history. Hosts are defined by the set of immune receptors they carry. 

We work in a 2-dimensional antigenic space, where each viral strain and each immune receptor in every host is a point in a 2D phenotypic space. This phenotypic space is motivated both by antigenic maps~\cite{Smith2004} and shape space used in immunology to describe the effective distance between immune receptors and antigen~\cite{perelson1979, Perelson1997, Chakraborty2010, Wang2015, Nourmohammad2016, Mayer2015a, Mayer2019}. The recognition probability of viruses by immune receptors is encoded in a cross-reactivity kernel $f(r)$ that depends on the distance between the virus and the receptor in this effective 2D space. We take $f(r) = e^{- {r}/{d}}$ to be an exponential function with parameter $d$, that determines the cross-reactivity --- the width of immune coverage given by a specific receptor~\cite{Luksza14}. 

All hosts start off with naive immune systems, implemented as a uniformly zero immune coverage in phenotypic space. If a host is infected by a virus, after the infection a new immune receptor is added to the host repertoire with a phenotypic position equivalent to the position of the infecting viral strain. Hosts have finite memory and the size of the memory pool of each host immune system $M$ determines the maximum number of receptors in a host repertoire, corresponding to the last $M$ viral strains that infected that host. This constraint can also be seen as the amount of resources that can be allocated to protect the host against that particular virus. 

An infection lasts a fixed time of $t_I= 3$ days, after which the infected host tries to infect a certain number of new hosts (among those who are not already infected), drawn from a Poisson distribution with average $R_0$. At this time the infection in the initial host is cleared and a memory immune receptor is added to its repertoire as explained above. During an infection a virus can mutate in the host with a rate $\mu$. Since we concentrate on the low mutation limit, ${\mu} t_I \ll 1$, we limit the number of per-host mutations to at most one. Following \cite{Bedford2012,Rouzine2018}, a mutation in a virus with phenotype $a$ produces a mutant with phenotype $b$ with probability density function $\rho(a \rightarrow b)=(1/2\pi)(4r_{ab}/\sigma^2) e^{-2r_{ab}/\sigma}$ (Gamma distribution of shape factor 2), where $r_{ab}$ is the Euclidean distance between $a$ and $b$, so that the average mutation effect is ${\sigma}$.
As a result the newly infected individual can be infected with the same (``wild-type'') virus that infected the previous individual with Poisson rate ${\rm e}^{- \mu t_I}$, or by a mutant virus with probability $P_{\rm mut} = 1 - {\rm e}^{- \mu t_I} $ for each infection event. 

Not all transmission attempts lead to an infection.  When a virus attempts to infect a host, an infection takes place with probability  $f(r)$, where $f$ is the cross-reactivity kernel defined above and $r$ is the distance in the 2D phenotypic space between the infecting viral strain and the closest receptor in the host repertoire. If the host repertoire is empty, the infection takes place with probability one. 
The viral mutation jump size and the cross-reactivity kernel set two length scales in the phenotypic space, $\sigma$ and $d$ (Fig.~\ref{cartoon}). Their dimensionless ratio $\sigma/d$ is one of the relevant parameters of the problem. In this work we kept $d$ fixed and then varied $\sigma$ to explore their ratio.
We do not explicitly consider competition between immune receptors within hosts, or complex in-host dynamics.

\subsection{Initial conditions and parameter fine-tuning}\label{inf_frac:sec}

We  simulate several such cycles of infections and recoveries, keeping track of the phenotypic evolution of viruses and immune receptors throughout time by recording the set of points describing viruses and receptors in phenotypic space at each time, as well as what immune receptors  correspond to each host. Once every 360 days we save a snapshot with the coordinates of all the circulating viruses. In addition we save the phylogenetic tree of a subsample of the viruses.  

In order to quickly reach a regime of co-evolution with a single viral lineage tracked by immune systems, we set initial conditions so that the viral population is slightly ahead of the population of immune memories. Details of the initial conditions are given in Appendix~\ref{sec:init}).

Viruses can survive for long times only because of an emergent feedback phenomenon that stabilizes the viral population when $R_0$ is fixed, as explained below in Section \ref{stability}. 
Even with that feedback, $R_0$ needs to be fined-tuned to obtain stable simulations.
With poorly tuned parameters, viruses go extinct very quickly after an endemic phase, as also noted in~\cite{Bedford2012}. The detailed procedure for setting $R_0$ is described in Appendix~\ref{inf_frac:sec_det}.
Roughly speaking, $R_0$ needs to be chosen so that the average effective number of infected people at each transmission event is equal 1, or $R_0p_f=1$, where $p_f$ is the average probability that each exposure leads to an infection. We further require that the fraction of infected hosts tends towards a target value, $\tilde f_{i}$, which acts as an additional parameter in our model. To do this, $R_0$ is first adaptively adjusted at each time as:
\beq\label{eq:R0}
R_0=\frac{1}{\<p_f\>}+\frac{\bar f_i-f_i}{\bar f_i},
\eeq
where $\<p_f\>$ is averaged over the past 1000 transmission events, and $f_i$ the current fraction of infected hosts. After that first equilibration stage, $R_0$ is frozen to its last value. Despite the explicit feedback ($\propto \bar f_i-f_i$) being removed, the population size is stabilized by the emergent feedback. 
As a result, the virus population is stable for long times for a wide range of parameter choices (Fig.~\ref{phylo_ph_diag_STAB}). 

To have more control on our evolution experiment we also analyze a variant of the model where we keep constraining the viral population size, constantly adjusting $R_0$ using Eq.~\ref{eq:R0} for the whole duration of the simulation (100 years). In this way the fraction of infected hosts $f_i$ is stabilized around the average $\bar f_i$.

Simulations were analyzed by grouping viral strains into lineages using a standard clustering algorithm, as described in Appendix~\ref{clust:sec}. The traces each lineage cluster were analyzed to evaluate their speed and variance in phenotypic space, as well as their angular persistence time (see Appendix~\ref{Appendix_pers} for details).
We built phylogenetic trees from subsamples of strains as detailed in Appendix~\ref{Appendix_phylo}.

\subsection{Detailed mutation model}\label{det_model}

We also considered a detailed in-host mutation model, in which we explicitly calculate the probability of producing a new mutant within a host. We present this model in detail in Appendix~\ref{one_mut_detailed_model} for the case where only one mutant reaches a high frequency during the infection time and we compare the results of this model to the simplified fixed mutation rate model described above.

The general idea is that we consider a population of viruses that replicate with rate $\alpha$ and mutate with  rate $\mu$ resulting in a non-homogeneous Poisson mutation rate $\mu  {\rm e}^{ \alpha t}$. The replication rate is the same for all mutants, i.e. there is no selection within one host and the relative fraction of the mutants depend only on the time at which the corresponding mutation arised.

For the case when only one mutation impacts the ancestral strain frequency, we simply calculate the time of the mutation event and use it to find the probability that an invader mutant  reaches a certain frequency at the end of the infection. We then randomly sample the ancestral or mutant strain according to their relative frequencies at the end of the infection to decide which one infects the next host.

\section{Results}

 % results

\begin{figure*}
\includegraphics[width=.7\linewidth]{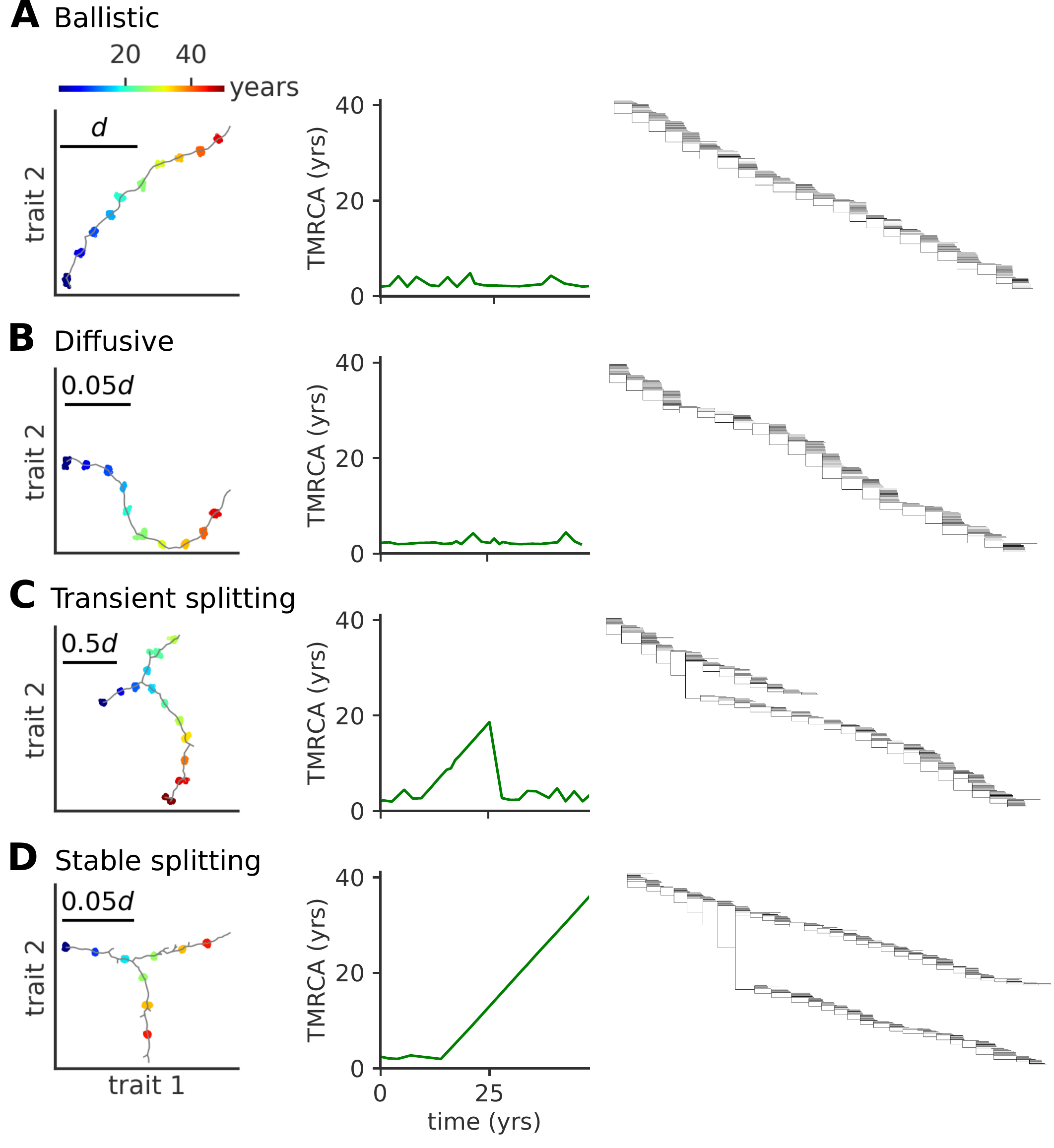}
\caption{\label{clust_traj}{\bf Modes of antigenic evolution: A) ballistic regime, B) diffusive regime, C) transient splitting regime, and D) stable splitting regime.} Left: trajectory of the population in phenotypic space (in units of $d$); Middle: time to most recent common ancestor (TMRCA); Right: phylogenetic tree of the population across time.
When viruses evolve in a single lineage the phylogenetic tree show a single trunk dominating evolution. When viruses split into more lineages, the phylogenetic tree shows different lineages evolving independently. 
Each lineage diffuses in phenotypic space with a persistence length that depends itself on the model parameters.  In these simulations viral population size is not constrained, but parameters are tuned to approach a target fraction of infected hosts, $\bar f_i = 10^{-3}$. Parameters are A) $\mu = 10^{-3}$,  $\sigma/d = 10^{-2} $, B) $\mu = 10^{-2}$,  $\sigma/d = 3\cdot 10^{-4}$ C) $\mu =  10^{-2}$,  $\sigma/d = 3\cdot 10^{-3}$,  D) $\mu = 0.1$,  $\sigma/d = 10^{-4}$. }
\end{figure*}

\subsection{Modes of antigenic evolution}

{Typical trajectories in phenotypic space show different patterns depending on the model parameters. In the following, we describe a ballistic (Fig.~\ref{clust_traj} A), a diffusive (Fig.~\ref{clust_traj} B), a transient splitting (Fig.~\ref{clust_traj} C), and a stable splitting (Fig.~\ref{clust_traj} D) regime and delineate the corresponding regions of the $\mu-\sigma$ parameter space.}

\noindent{\em Ballistic regime.}
In this regime of one-dimensional evolution, viruses mutate locally forming a concentrated cluster of similar individuals, called a lineage. Successful mutation events that take the viral strains away from the protection of hosts immune systems progressively move the lineage forward (Fig.~\ref{clust_traj} A). For small values of the mutation rate and small mutation jump sizes the trajectory in phenotypic space is essentially linear, with new mutants always growing as far away as possible from existing hosts immune system, which themselves track viruses but with a delay. The delayed immune pressure creates a fitness gradient for the virus population, which forms a traveling fitness wave \cite{Desai2007a, Neher2013,Yan2018} fuelled by this gradient. A similar linear wave scenario was studied in one dimension by Rouzine and Rozhnova \cite{Rouzine2018}.

\noindent{\em Diffusive regime.}
As we increase the mutation jump range the trajectories loose their persistence length and the trajectories in phenotypic space start to turn randomly, as new strains are less sensitive to the pressure of hosts immune systems (Fig.~\ref{clust_traj} B).

Both ballistic and diffusive regimes lead to phylogenetic trees with one main trunk and a short distance to the last common ancestor. This trend is characteristic of influenza A evolution and has been discussed in detail in Ref.~\cite{Bedford2012}.

\noindent{\em Transient splitting regime.}
Alternatively, we observe a bifurcation regime, where at a certain point in time two mutants form two new co-existing branches, roughly equidistant from each other and the ancestral strain in antigenic space (Fig.~\ref{clust_traj} C). Each branch has similar characteristics as the single lineage in the one dimensional evolution of Fig.~\ref{clust_traj} A and B. These co-existing branches give rise to phylogenetic trees with two trunks. In the example shown the two lineages stably co-exist for $\sim 20$ years, leading to a linear increase of the distance to the last common ancestor, until one of them goes extinct, returning the evolution to one dominant lineage with small distances to the last common ancestor. 
 
\noindent{\em Stable splitting regime.}
The two branches can stably co-exist for over $\sim 80$ years (Fig.~\ref{clust_traj} D, only the first 50 years are shown), starting with similar trends as in the example in Fig.~\ref{clust_traj} C, not returning to the one dominant lineage regime, but even further branching in a similar equidistant way at later times (not shown). This trend leads to evolutionary trees with multiple stable trunks, with local diversity within each of them and a linear increase of the distance to the last common ancestor over long times.

\begin{figure}
\includegraphics[width=0.75\linewidth]{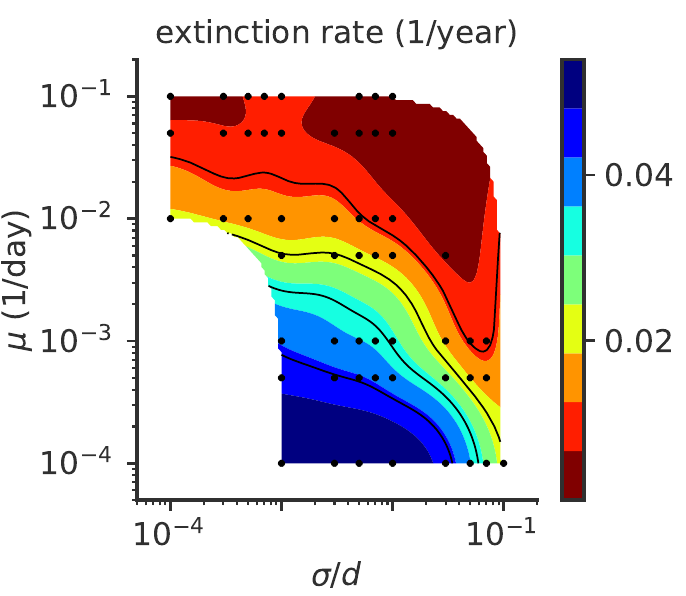}
\caption{\label{phylo_ph_diag_STAB}{\bf The rate at which viruses go extinct depends on model parameters.} Viruses extinction rate (1/year) as a function of $\mu$ and $\sigma$.  In these simulations viral population size is not constrained, and $\bar f_i = 10^{-3}$. For each parameter point we simulated 100 independent realizations.}
\end{figure}

\subsection{Stability}~\label{stability}

The extinction rate of viral populations depends on the parameter regime (Fig.~\ref{phylo_ph_diag_STAB}). A stable viral population is achieved in the $\sigma \ll d $ regime thanks to stabilizing feedback~\cite{Yan2018}: if viruses become too abundant they drag the immune coverage onto the whole viral population, and the number of viruses decreases since infecting a new host becomes harder. As a result the relative advantage of the fittest strains with respect  to the bulk of the population decreases as more hosts are protected against all viruses. This feedback slows down the escape of viruses to new regions of antigenic space and the adaptation process. Conversely, when the virus abundance drops, the population immune coverage is slower in catching up with the propagating viruses. The fittest viral strains gain a larger advantage with respect to the bulk  and this drives viral evolution faster towards new antigenic regions and higher fitness, increasing the number of viruses. 

This stabilizing feedback is very sensitive to the speed and amplitude of variation. Abrupt changes or big fluctuations in population size can drive the viral population to extinction. Because of this, viruses often go extinct very quickly after an endemic phase~\cite{Bedford2012, Yan2018}, as is proposed to have been the fate of the Zika epidemic~\cite{Yan2018}. Here we focus on the stable evolutionary regimes, starting from a well equilibrated initial condition as explained in section~\ref{inf_frac:sec}. 

\begin{figure*}
\includegraphics[width=.8\linewidth]{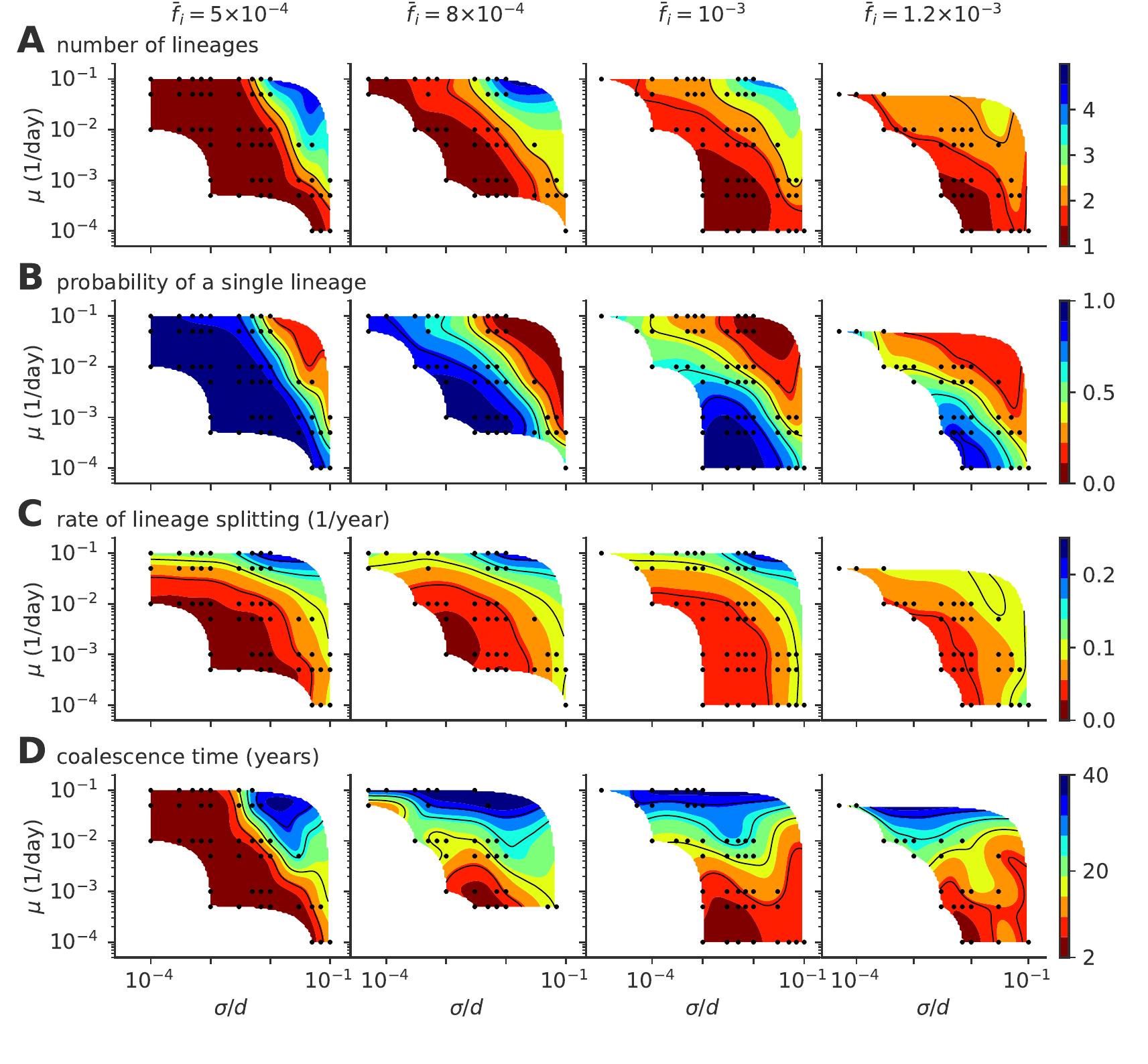}~\label{pheno_ph_diag}
\caption{\label{pheno_ph_diag}{\bf Phase diagram of the single- to multiple lineage transition as a function of mutation rate $\mu$, mutation jump size $\sigma$, and  $\bar f_i$.} (A) Average number of lineages, (B) fraction of evolution time where viruses are organized in a single lineage, (C) rate of lineages splitting (per lineage), and (D) average coalescence time. In these simulations viral population size is not constrained, and the target fraction of infected individuals $\bar f_i$  is $5\cdot 10^{-4}$, $8\cdot 10^{-4}$, $ 10^{-3}$, $1.2\cdot 10^{-3}$, from left to right. For each parameter point we simulated 100 independent realizations.}
\end{figure*}

\subsection{Phase diagram of evolutionary regimes}

Our results depend on three parameters: the mutation rate $\mu$, the mutation jump distance measured in units of cross-reactivity $\sigma/d$, and the target fraction of infected individuals in the population, $\bar f_i$. The observed evolutionary regimes described in  Fig.~\ref{clust_traj} depend on the parameter regimes, as summarized in the phase diagrams presented in Fig.~\ref{pheno_ph_diag} 
for various fractions of infected hosts $\bar f_i$.

The mean number of distinct stable lineages increases with both the mutation rate and the mutation jump distance (Fig.~\ref{pheno_ph_diag} A). Because the process is stochastic, even in regimes where multiple lineages are possible, particular realizations of the process taken at particular times may have one or more lineages.
The fraction of time when the population is made of a single lineage decreases with mutation rate and jumping distance (Figure~\ref{pheno_ph_diag} B), while the rate of formation of new lineages increases (Fig.~\ref{pheno_ph_diag}C). All three quantities indicate that large and frequent mutations promote the emergence of multiple lineages.
This multiplicity of lineages arises when mutations are frequent and large enough so that two simulataneous escape mutants may reach phenotypic positions that are distant enough from each other so that their sub-lineages stop feeling each other's competition and become independent.

Increasing the mutation rate or the mutation jump distance alone is not always enough to create a multiplicity of lineages. For small $\bar f_i=5\cdot 10^{-4}$ and moderate jump sizes, the single-lineage regime is very robust to a large increase in the mutation rate, meaning the cross-immunity nips in the bud any attempt to sprout a new lineage from mutations with small effects, however frequent they are.

Coalescence times (Fig.~\ref{pheno_ph_diag} D) give a measure of the number of mutations to the last common ancestor, and are commonly used in population genetics to characterize the evolutionary dynamics. In the case of a single lineage, coalescence times are short, corresponding to the time it takes for an escape mutation furthest away from the immune pressure to get established in the population. However, when there are multiple lineages, the coalescence time corresponds to the last time a single lineage was present. Such an event can be very rare when the average nmber of lineages is high, leading to very large coalescence times. Accordingly, the coalescence time increases with lineage multiplicity, and thus with mutation rate and jump size.

In general, large target fraction of infected hosts, $\bar f_i$, lead to more lineages on average and a higher probability to have more than one lineage.
Increasing the number of infected individuals  increases the effective mutation rate and allows the virus to explore evolutionary space faster. This rescaling allows more viruses to find niches and increases the chances of having co-existing lineages.  While an increased fraction of infected hosts may also limit the virgin exploration space where viruses can attack non-protected individuals, this effect may be negligible when the target fractions $\bar f_i$ are small as considered here.

\begin{figure}
\includegraphics[width=\linewidth]{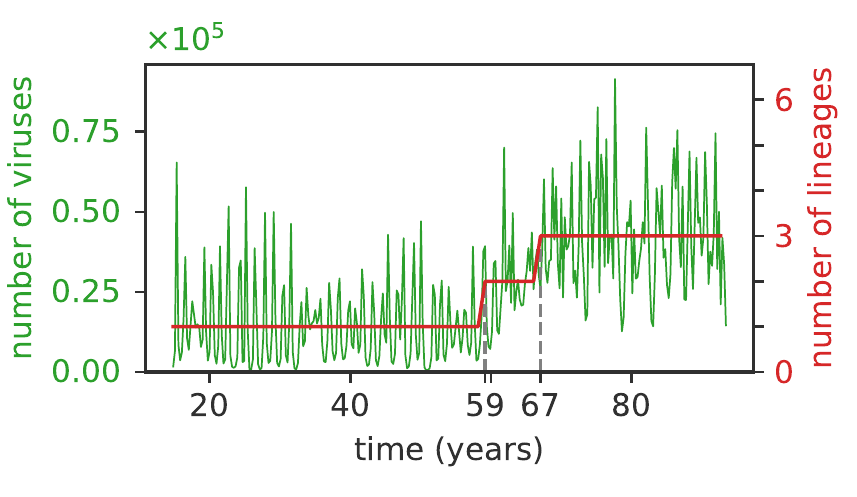}
\caption{\label{vir_num_per_cl}{\bf The average number of viruses is proportional to the number of independent clusters.} The total number of viruses (green curve) and of lineages (red curve) as a function of time  for $\bar f_i = 10^{-3}$, $\mu = 10^{-2}$,  $\sigma/d = 3\cdot 10^{-3}$. The initial single lineage splits into two lineages at $t\approx 59$ years and then  into three lineages at $t\approx 67$ years (dashed vertical lines), and the number of viruses first doubles and then triplicates following the lineage splittings.}
\end{figure}

\subsection{Incidence rate}

When viruses split into lineages, the implicit feedback mechanism described earlier to explain stability remains valid for each cluster independently (unless the number of independent lineages exceeds the immune memory pool $M$). As a result each lineage can support roughly a fraction $\bar f_i$ of the hosts, which defines a ``carrying capacity'' of each lineage.
As a result the viral population size, also known as incidence rate, is proportional to the number of lineages (Fig.~\ref{vir_num_per_cl}).
Yet the incidence fluctuates with time, with clear bottlenecks when a new cluster is founded. 

\begin{figure*}
\includegraphics[width=.8\linewidth]{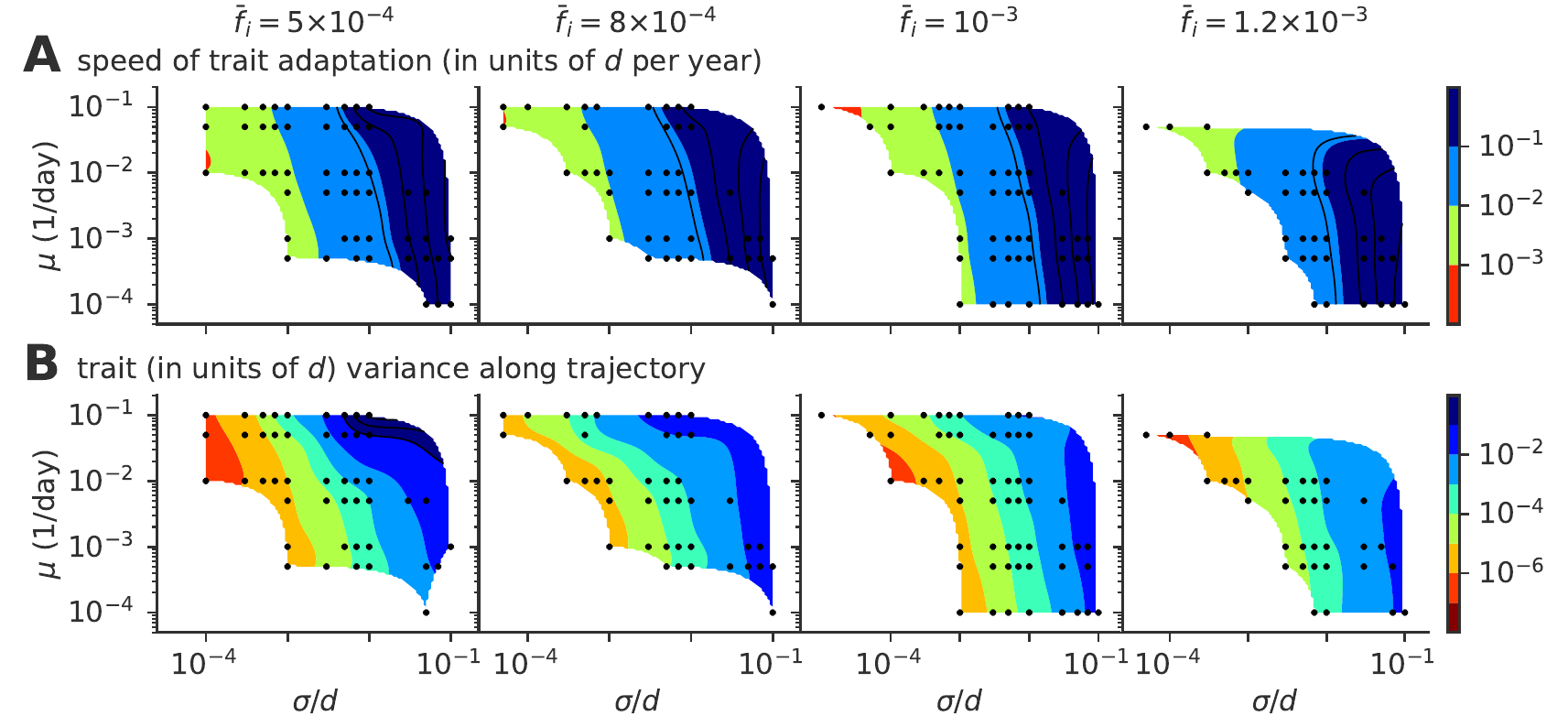}
\caption{\label{var_vel}{\bf Speed of adaptation and the within-cluster diversity.} Phase diagrams as a function of mutation rate $\mu$ and mutation jump rate $\sigma$ for  (A) the average speed of the evolving viral lineages and (B) the variance of the size of the cluster in the direction parallel to the direction of instantaneous mean adaptation for different values of target infected fraction $\bar f_i=5\cdot 10^{-4}$, $8\cdot 10^{-4}$, $10^{-3}$, and $1.2\cdot 10^{-3}$ from left to right. For each parameter point we simulated 100 independent realizations.}
\end{figure*}

\subsection{Speed of adaptation and intra-lineage diversity}

Whether there is a single lineage or multiple ones, each lineage moves forward in phenotypic space by escaping the immune pressure of recently infected and protected hosts lying close behind.
We examined  the speed of adaptation and the diversity of lineages of viral diversity present at a given time (Fig.~\ref{var_vel}). We calculated the speed of adaptation in units of cross-reactivity radii $d$ per year by taking, for each lineage, the difference in the two dimensional phenotypic coordinate of the average virus at time points one year apart. 
We quantified the diversity by approximating the density of each lineage at a given time by a Gaussian distribution in two-dimensional phenotypic space and calculating its variance along the direction of the lineage adaptation in phenotypic space.

The speed of adaptation increases with the mutational jump size $\sigma$, and also shows a weak dependence on the mutation rate $\mu$. 
The variance in the viral population also increases with the jump size, and in general scales with the speed of adaptation. Fisher's theorem states that the speed of adaptation is proportional to the fitness variance of the population. A correspondance between speed and variance in phenotypic space is thus expected if fitness is linearly related to phenotypic position. While such a linear mapping does not hold in general in our model, the immune pressure does create a nonlinear and noisy fitness gradient, which can explain this scaling between speed and diversity.

\begin{figure*}
\includegraphics[width=.8\linewidth]{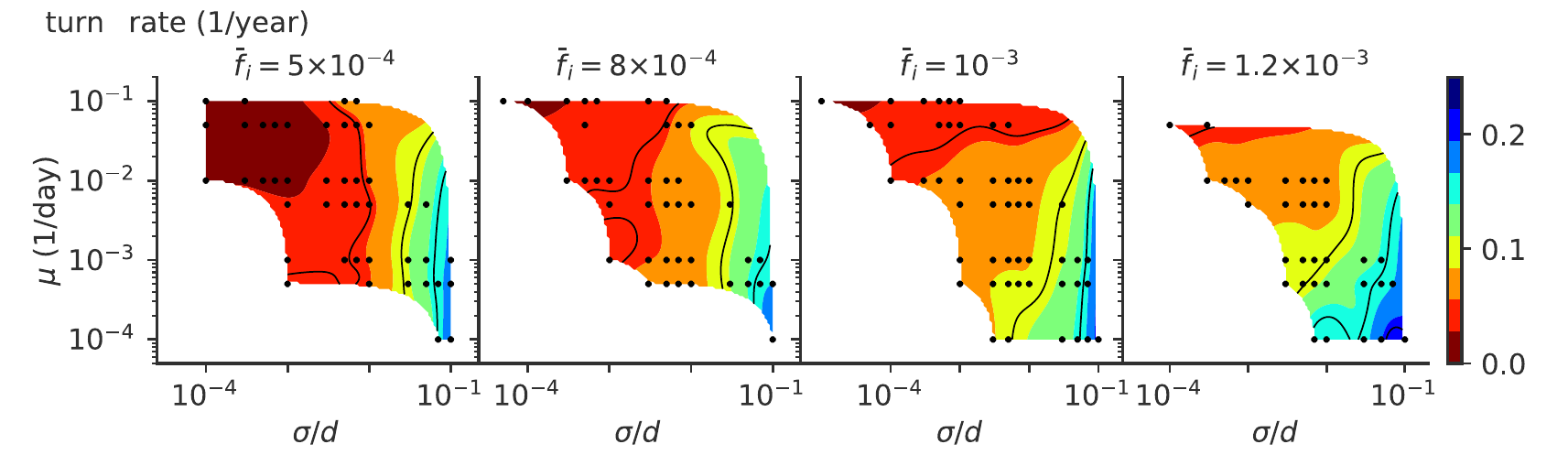}
\caption{\label{pers}{\bf Turn rate.} Phase diagrams as a function of mutation rate $\mu$ and mutation jump rate $\sigma$ for rate of turns (defined as a change of direction of at least 30 degrees) of the trajectories, for different values of the mean number of infected individuals $\bar f_i$: $5\cdot 10^{-4}$, $8\cdot 10^{-4}$, $10^{-3}$, $1.2\cdot 10^{-3}$, from left to right.}
\end{figure*}

\subsection{Antigenic persistence}
While lineage clusters tend to follow a straight line, their direction fluctuates as escape mutants can explore directions that are orthogonal to the main direction of the immune pressure.
In Fig.~\ref{pers} we plot the rate at which trajectories turn, changing their direction by at least 30 degrees (see Appendix~\ref{Appendix_pers}). As noted in Fig.~\ref{clust_traj}, small mutation jump sizes $\sigma$ favor long periods of linear motion and low turn rates.  As $\sigma$ increases, the turn rate increases.

Several factors affect the turn rate as measured from the simulations.
A lineage splitting induces a turn, and regions of phase space where multiple lineages are possible favor short persistence times. The same goes for population extinction: regimes where the population extinction rate is higher do not allow us to observe long persistence times, masking the dependence of turn rate on $\mu$. 
Generally, we expect lineage clusters to undergo more angular diffusion in phenotypic space as mutations become more important (large $\sigma$). Mutants can explore new regions of the phenotypic space, causing the population to stochastically turn while keeping a cohesive shape. On the other hand, lower mutation rates may mean that fewer mutants will do this exploration, increasing stochasticity in cluster dynamics and effectively increasing the turn rate.
In that regime of stochastic turning, predictions of the phenotype of future viral strains is much harder than in the linear regime.

\begin{figure}
\includegraphics[width=\linewidth]{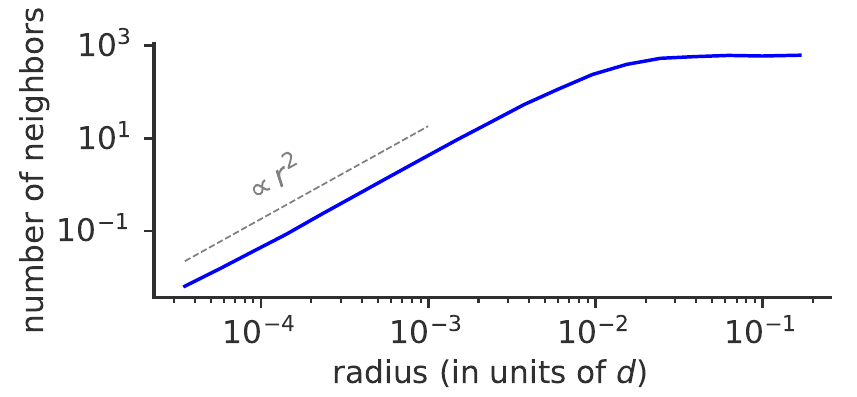}
\caption{\label{Michael1}{\bf Effect of phenotypic space dimensionality on viral evolution.} Cumulative average number of neighbours of a given viral strain as a
function of phenotypic distance $r$ to that strain  for $\bar f_i =10^{-3}$, $\mu = 10^{-2}$,  $\sigma/d = 3\cdot 10^{-3}$.  The  average number of neighbours depends on the dimension of the phenotypic space as $r^{D}$ where $r$ is the distance and $D=2$ the dimension of phenotypic space (dotted line).}
\end{figure}

\subsection{Dimension of phenotypic space}
We explored the role of phenotypic space dimensions in our results. In Fig.~\ref{Michael1} we plot the average number of neighbours of a given viral strain within distance $r$ from that strain (for short distances so that only pairs from the same lineage are considered). This  measure scales as $r^D$ for the cumulative number of neighbours plotted in Fig.~\ref{Michael1}, where $D=2$ the dimension of phenotypic space, as expected for a uniformly distributed cluster of strains in finite dimension. By contrast, that number would be expected to scale exponentially with $r$ for a neutral process in infinite dimensions. This results suggests that in low dimensions, which seems to be the experimentally valid limit, the dimension of the space does restrict the dynamics and cannot be neglected. However we are unable to separate the effects of selection and phenotypic space dimensionality. It also implies that lineages form dense, space-filling clusters in phenotypic space. We expect this result to hold for any reasonably low dimension, and will break down in high dimension.

\begin{figure}
\includegraphics[width=\linewidth]{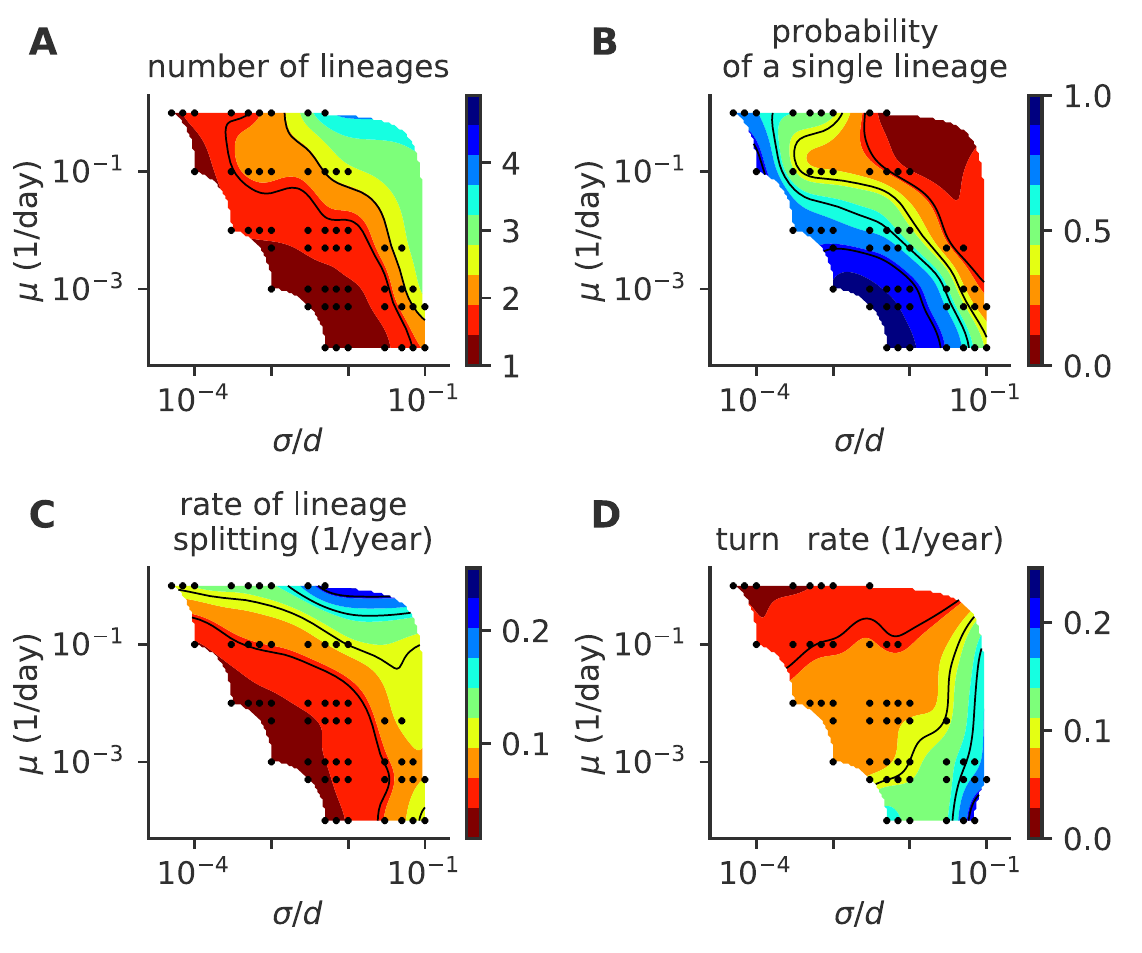}
\caption{\label{det_mut}{\bf Phase diagram for the detailed intra-host mutation model}. As a function of the mutation rate $\mu$ and mutation jump size $\sigma$ we plot (A) the mean number of co-existing lineages, (B) the fraction of time with one lineage,  (C) the lineage splitting rate and  (D)  the lineage turn rate.  In these simulations viral population size is not constrained, and $\bar f_i =10^{-3}$. For each parameter point we simulated 100 independent realizations.}
\end{figure}

\begin{figure}
\includegraphics[width=\linewidth]{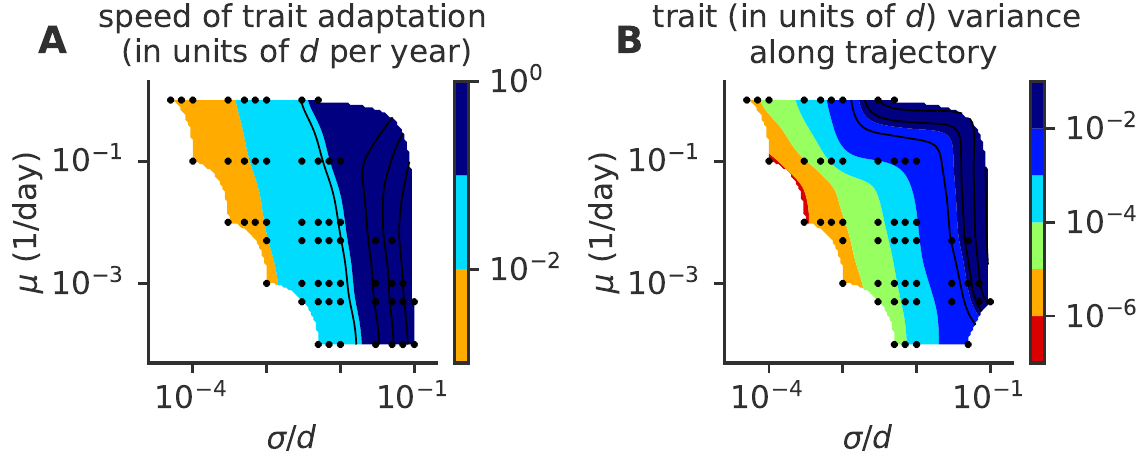}
\caption{\label{det_mut_vel}{\bf Phase diagram for speed of adaptation and within cluster diversity of the detailed intra-host mutation model}. As a function of the mutation rate $\mu$ and mutation jump size $\sigma$ we plot (A) the mean  speed of adaptation, and (B) the variance in the cluster size in the direction parallel to the direction of motion.  In these simulations viral population size is not constrained, and $\bar f_i =10^{-3}$. For each parameter point we simulated 100 independent realizations.}
\end{figure}

\subsection{Robustness to details of intra-host dynamics and population size control}~\label{res:det_mutation_model}

To test whether a detailed treatment of intra-host viral dynamics would affect our results,
we also considered a detailed mutation model, where we calculate the probability of  producing a mutation within in each individual (see Appendix~\ref{one_mut_detailed_model}). Specifically, we compare the model that calculates the probability of having a mutated strain  given in Eq.~\ref{P_invader_res} to the simplified model with simple mutation rate discussed above.  As we see from Fig.~\ref{det_mut} and  Fig.~\ref{det_mut_vel}, the general evolutionary features are the same as for the simplified model: the probability of multi-lineage trajectories increases with increasing $\mu$ and $\sigma$, as does the lineage splitting rate and the speed of adaptation. The diversity in phenotypic space in the direction parallel (Fig.~\ref{det_mut_vel} B) to the direction of motion
increases with the mutation jump size, as expected, as well as the turn rate (Fig.~\ref{det_mut} D).

Lastly we asked how our results would be affected by strictly constraining the viral population size (as explained in Sec.~\ref{inf_frac:sec}), rather than letting it fluctuate under the control of the emergent negative feedback. The corresponding phase diagrams show the same evolutionary regimes as a function of $\mu$ and $\sigma/d$ (Fig.~\ref{pheno_ph_diagFIX}), and the same general dependencies on model parameters of the speed of adaptation (Fig.~\ref{var_velFIX}) and turn rate (Fig.~\ref{pers_fix}), as with a fluctuating population.

\section{Discussion}

 % discussion

{Our model describes regimes of viral evolution with different complexity: one strain dominates (Fig.~\ref{clust_traj} A, B), two dominant strains coexist over timescales longer than the host lifetime (Fig.~\ref{clust_traj} C), or multiple strains coexist in a stable way (Fig.~\ref{clust_traj} D). The single-strain regime clearly maps onto influenza A. Influenza B evolution, which is split into the Victoria or Yamagata sublineages, is consistent with prolonged (Fig.~\ref{clust_traj} C) or stable coexistence (Fig.~\ref{clust_traj} D). We can} use our results to characterize the differences in the evolutionary constraints acting on the adaptive processes of influenza A and B. Our results suggest that the combination of mutation rate and effective mutation jump distance in influenza A must be smaller than in influenza B. Since the mutation rates are similar, this means that the effect of mutations at sites in influenza B has a larger different phenotypic affect. Alternatively, the effective number of infected individuals per transmission event ($R_0$ in classical SIR models, equal to $R_0p_f$ in our model) could be larger in influenza B than influenza A. Another possibility is that, since lineage splitting happens stochastically, the difference between the two species is just due to different random realizations. 

Our model shares similarities with previously considered models of viral evolution~\cite{Bedford2012, Yan2018}, while focusing on distinct questions. Among differences in modeling details,
our hosts have finite memory capacity and forget past strains after some time, compared to infinite memory assumed in past work. Comparing our simulations with Ref.~\cite{Bedford2012, Yan2018} in their relevant regimes, we do not see noticeable differences in the main trends of evolution, which suggests that the effects of losing memory are quantitative rather than qualitative at the population scale, at least for the parameters regimes that were inspected. We assume exponentially decaying cross-reactivity, similarly to \cite{Yan2018} (although it is linearized in their analysis). By contrast, Ref.~\cite{Bedford2012} uses a linear cross-reactivity, but this minor difference is unlikely to influence the results. Ref.~\cite{Bedford2012} focused specifically on the question of explaining the single dominant lineage in influenza A evolution. While the existence of lineage bifurcations was acknowledged in Ref.~\cite{Bedford2012}, this regime was not explored. Instead, a more detailed geographical model was considered, with migrations between different geographical zones. Ref.~\cite{Yan2018} asked a similar question we did about the conditions under which strain bifurcations may occur, but in the context of an infinite antigenic space. The general trends seem to be independent of the dimensionality of the space, since both models recover the same behaviour. However, the exact scaling laws reported in Ref.~\cite{Yan2018} seem to be more sensitive to the model assumptions. Lastly, while we also considered a more detailed model of intra-host influenza evolution, we found that it could be mapped onto an effective model of viral transmission with mutations, with little impact on the results.

Two main effective parameters control the evolutionary patterns: the effective mutation rate and the mutational jump size, measured in units of the cross-reactivity radius. The effective mutation rate is a combination of the actual mutation rate per host, and the mean number of infected hosts at each cycle: larger fractions of  infected individuals lead to more opportunities for the virus to escape host immunity, and faster viral adaptation as a whole. Additionally, a feedback mechanism is observed between the host immune systems and the viruses: too successful viruses infect many hosts, effectively speeding up the rate at which the susceptible host reservoir is depleted, and mounting up the immune pressure. Our model does not include host death, since we assume we are in the limit of very large host reservoirs.
Accounting for host extinction may leads to a different interesting problem that has been explored using SIR models~\cite{Grenfell2002, Allen2012}. In the context of our model however, host death would effectively amount to reducing the hosts' immune memory capacity $M$.

The effects of dimensionality on the observed evolutionary trajectories are worth discussing in more detail. The infinite dimensional model is similar in spirit to the infinite sites model of sequence evolution: infinite dimensions mean there is always a direction for the virus to escape to. Conversely, low dimensions result in an effectively stronger feedback of the host immune systems on the possible trajectories of the escaping virus. {This generates effective mutation and jump rates that depend on the primary parameters in a nonlinear way, with possibly different effects in different parameter regimes.}
We also observe a breakdown of the scaling of observables such as the coalescence time and the mean number of co-existing lineages with $\mu \sigma^2$ (see Fig.~\ref{fig:scaling}), as would be predicted by the diffusion limit of the traveling wave framework~\cite{Cohen2005,Neher2013}. These results indicate that the discreteness of mutations matter. The effective dimensionality of the phenotypic space depends on the parameters, going from effectively one in the linear regime to the dimension of the phenotypic space in the splitting regime. We expect that our results generalize to higher dimensions than 2, with each splitting event leading to a new direction in phenotypic space and increasing the effective dimension of the viral population.

In summary, a detailed exploration of the mutation rate and jump distance, as well as the fraction of infected individuals allowed us to understand the constraints that lead to different {modes of antigenic evolution and, in particular, lineage splitting at different rates and with different survival times of new (sub-)lineages.  Observed bifurcations are rare in nature, which puts an evolutionary constraint on the adaptation process. 
~\label{discussion}

{\bf Acknowledgements. 
} This work was partly supported by ERC CoG 724208 and DFG grant CRC 1310 "Predictability in Evolution". 
\medskip

\bibliographystyle{pnas}

\appendix

 % appendixA
\section{Simulation details}

\subsection{Initialization}\label{sec:init}
We initialize all simulations in an immune coverage background that favors the evolution of one dominant antigenic lineage. We draw viral positions uniformly in a rectangle with bottom-left and top-right corners positioned at  $(- 3 \sigma P_{\rm mut},0)$ and $( 3 \sigma P_{\rm mut}, \sigma)$. Each host is initialized with one immune receptor as a point in antigenic space, which grants localized protection. The initial memory repertoires of the different hosts are drawn uniformly from a rectangle with bottom-left and top-right corners positioned at $(- 3 \sigma P_{\rm mut},- 5 \frac{ \sigma}{\bar f_i} P_{\rm mut})$ and $( 3 \sigma P_{\rm mut}, 0)$, where $\bar f_i$ is the target fraction of infected hosts, determining the number around which the viral population is stabilized (see Section~\ref{inf_frac:sec}) and the timescale with which all hosts add (or renew) an immune receptor to their repertoire. In order to lose memory of the artificial initial conditions we let the system evolve until $99\%$ of the host population have been infected by a virus, so that most hosts have added at least one strain to their repertoires before recording any data.

\subsection{Control of the number of infected hosts}\label{inf_frac:sec_det}

We studied two versions of the same model, one constraining the viral population size strictly, the other letting it fluctuate. In the latter case, we still have to constrain population size for an initial transient in order to reach a well equilibrated initial condition.

We control the virus population size through the fraction of infected hosts around a target value of $\bar f_i$. We modify $R_0$ --  the average number of new hosts that are drawn to be infected in a given transmission event -- based on the current fraction of infected hosts $f_i$ at each time:
\begin{equation}\label{R_0}
R_0 =\frac{1}{\langle p_f \rangle} + \frac{\bar f_i - f_i}{\bar f_i}  \,,
\end{equation}
where $p_f$ is the probability of a successful infection at a transmission event, {i.e.} the probability that a new host is susceptible to the infecting viral strain. We evaluate its average $\langle p_f \rangle$ over segments of 1000 transmission events.

On average,
\begin{equation}\label{f_i_recursive}
\<f_i(t+t_I)\> \approx \<f_i(t)\> R_0 \<p_f\>.
\end{equation}
Using eq.~\eqref{R_0}, we find that the average fraction of infected hosts $\<f_i(t)\>$ is governed by a logistic map with fix point $\bar f_i$,
effectively producing a process where the viral population growth is limited by an effective carrying capacity $N\bar f_i$.

 % appendixB
\section{Detailed mutation model}~\label{one_mut_detailed_model}

We present the detailed in-host mutation model, in which we explicitly find  the probability of producing a new mutant within an infected host. We assume that the immune system responds only to the first viral strain it sees, and  that all viruses see the immune system in the same way, undergoing the same deterministic dynamics, i.e. evolution is neutral within one host.
This intra-host neutral selection holds if the characteristic mutation jump size is smaller than the cross reactivity length, $\sigma \ll  d$, which is the case for our simulations. We consider this mutation-proliferation process up to time $t_I$. 

We call the total viral population $v_{tot}$, the first viral invader, that is the first viral strain infecting one host, $v_{0}$, and the new mutants, appearing with size 1, $v_{j}$. These three quantities (neglecting the discreteness of the process) grow deterministically as function of time $t$ as:
\begin{align}\label{gr_laws_mutants}
v_{tot}(t) &= {e}^{\alpha t} \,, \\
v_{0}(t) &= {e}^{\alpha t} - \sum_{i_{0}} {e}^{\alpha (t - t_{i_{0}})}\Theta(t- t_{i_{0}})\,, \\
v_{j}(t) &= {e}^{\alpha (t - t_{j})} - \sum_{i_{j}} {e}^{\alpha (t - t_{i_{j}})}\Theta(t- t_{i_{j}})\,,
\end{align}
where $i_{j}$ denotes the indexes of the viral mutants originated from mutant $j$ (if any) and $t_{i_{j}}$ indicates the times at which such mutations arose ($\Theta(x)$ is the Heaviside function, $=0$ for $x<0$ and 1 otherwise). Each mutation jumps to new phenotypic coordinates. From these equations the relative mutants fractions are 
\begin{align}\label{fracts_mutants}
x_{0}&= 1 - \sum_{i_{0}} {e}^{- \alpha t_{i_{0}}}\Theta(t- t_{i_{0}})\,, \\
x_{j} &= {e}^{-\alpha t_{j}} - \sum_{i_{j}} {e}^{- \alpha t_{i_{j}}}\Theta(t- t_{i_{j}}).
\end{align}

The mutation process from any virus present in the viral pool is a non homogeneous Poisson process with rate $\mu  {e}^{\alpha t}$. The probability of having $n$ mutations up to the time $t$ is:
\begin{equation}\label{pNt_Poiss_nonhomo}
P(n, t) = \frac{(\Lambda(t))^n}{n!} {e}^{-\Lambda(t)} \,,
\end{equation}
with 
\begin{equation}\label{Lambda_Poiss_nonhomo}
\Lambda(t) = \int_0^t dt' \mu e^{\alpha t'}=\frac{\mu}{\alpha} ({e}^{\alpha t}- 1) \,.
\end{equation}
The time $t_1$ of the first mutation event is distributed as:
\begin{equation}\label{1st_Poiss_nonhomo}
\rho(t_1) = \mu {e}^{\alpha t_1-\Lambda(t_1)}.
\end{equation}

In our simulations, we assume that all mutations other than the first are negligible, that is, we can have more than one mutation, but those after the first don't affect significantly the relative fraction, therefore we have only one mutant. The fraction of the mutant is $x_{1}(t) = {e}^{- \alpha t_{1}}$ if $t >t_1$. Knowing the distribution of the first mutation times $t_1$, we can calculate the probability distribution of the mutant fraction $x_{1}$ at the time of the transmission event $t_I$:
\begin{equation}\label{P_invader_res}
\rho(x_{1}, t_I) = {e}^{-\Lambda(t_I)} \delta(x_{1}) + \frac{\mu  {e}^{- \frac{\mu}{\alpha} (\frac{1}{x_{1}} -1)} }{\alpha x_{1}^2} \Theta(x_1-e^{-\alpha t_I})\,.
\end{equation}

In the simulations we fixed the growth rate to $\alpha=4$ day${}^{-1}$.

 % Appendix_pers
\section{Analysis of simulations}\label{analysis}

\subsection{Lineage identification}\label{clust:sec}

In order to analyze the organization of viruses in the phenotypic space, for each saved snapshot we take the positions of a subset of 2000 viruses and then cluster them into separate lineages through the python scikit-learn DBSCAN algorithm~\cite{scikitlearn}~\cite{dbscan} 
with the minimal number of samples parameter $min\_samples = 10$. We perform the clustering for different values of the parameter $\epsilon$ defining the maximum distance between two samples for one to be considered as in the neighborhood of the other, and then select the value that minimizes the variance of the 10th nearest neighbor distance (the clustering results are not sensitive to this choice). From the clustered lineages we can easily obtain a series of related observables, such as the number of lineages and the fraction of time in which viruses are clustered in a single lineage (Fig.~\ref{pheno_ph_diag}). A split of a lineage into two new lineages is defined when  two clusters are detected where previously there was one, and the two new clusters centroids are farther away than the sum of the maximum distances of all the points in each cluster from the corresponding centroid. We impose this extra requirement in order to reduce the noise from virus subsampling and the clustering algorithm. A cluster extinction is defined when a cluster ceases to be detected from one snapshot to the next. 

\subsection{Turn rate estimation}\label{Appendix_pers}
We estimate the turn rate 
by detecting turns in the trajectories of lineages centroids in phenotypic space. This is done by calculating the trajectory's angle between subsequent centroids recordings and smoothing it with a 5 year averaging window. A turn is detected when the angle difference with respect to the initial direction reaches 30 degrees, and the time before the turn is recorded as the persistence time. Then the procedure is repeated until the end of the trajectory. 
In order to have enough timepoints in the trajectory, we limit this analysis to lineages that last more than 20 years.  This is done for all the lineages trajectories in all the realization.
Finally to estimate the turn rate we divide the total number of detected turns by the sum of the durations of all the analyzed trajectories. 

\subsection{Phylogenetic tree analysis}\label{Appendix_phylo}
From the model simulations we record a subsample of the viral phylogenetic tree. For every recorded strain, apart from some descendants we also save their extinction events.  To compute the coalescence time we take the recorded circulating strains  once every year, that is all the strains recorded before that year that have not gone extinct yet. Then we calculate the time to their most recent common ancestor, and finally we average over all these TMRCAs calculated year after year, for all the realizations. 
Phylogenetic tree analysis and rendering are done using the python open software ETE Toolkit~\cite{ETE}.

\beginsupplement

 % Appendix

\begin{figure*}
\includegraphics[width=.7\linewidth]{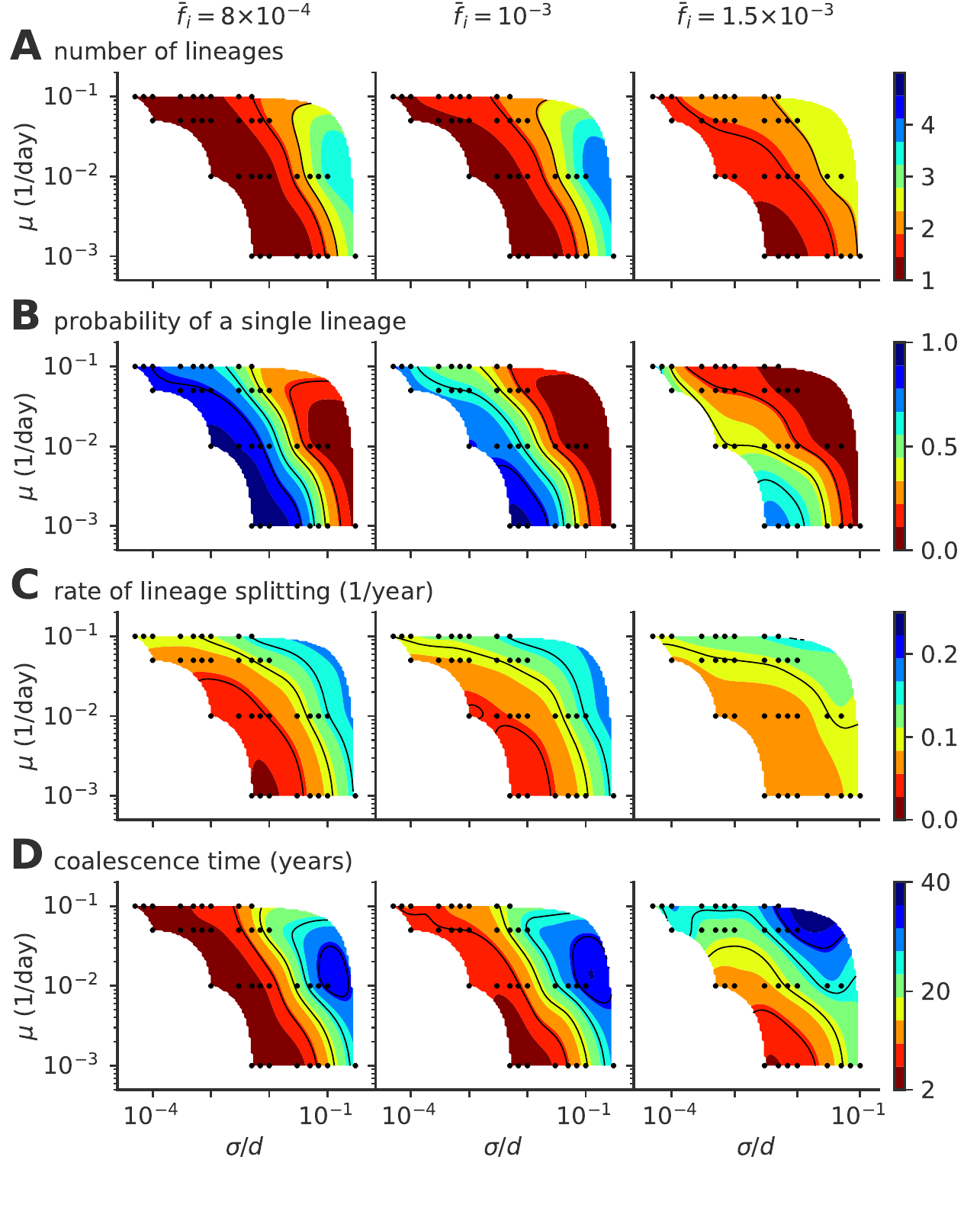}
\caption{\label{pheno_ph_diagFIX}{\bf Phase diagram for a fixed population size of the single- to multiple lineage transition, as a function of mutation rate $\mu$ and mutation jump size $\sigma$.} The figure is similar to the one presented in the main text in Fig.~\ref{pheno_ph_diag} but assuming a fixed fraction of infected hosts $\bar f_i=8\cdot 10^{-4}$, $10^{-3}$, and $1.5\cdot 10^{-3}$ (from left to right). (A) Average number of lineages, (B) fraction of time where viruses are organized in a single lineage, (C) rate of lineage splitting, and (D) taverage coalescence time (D).}
\end{figure*}

\begin{figure*}
\includegraphics[width=.7\linewidth]{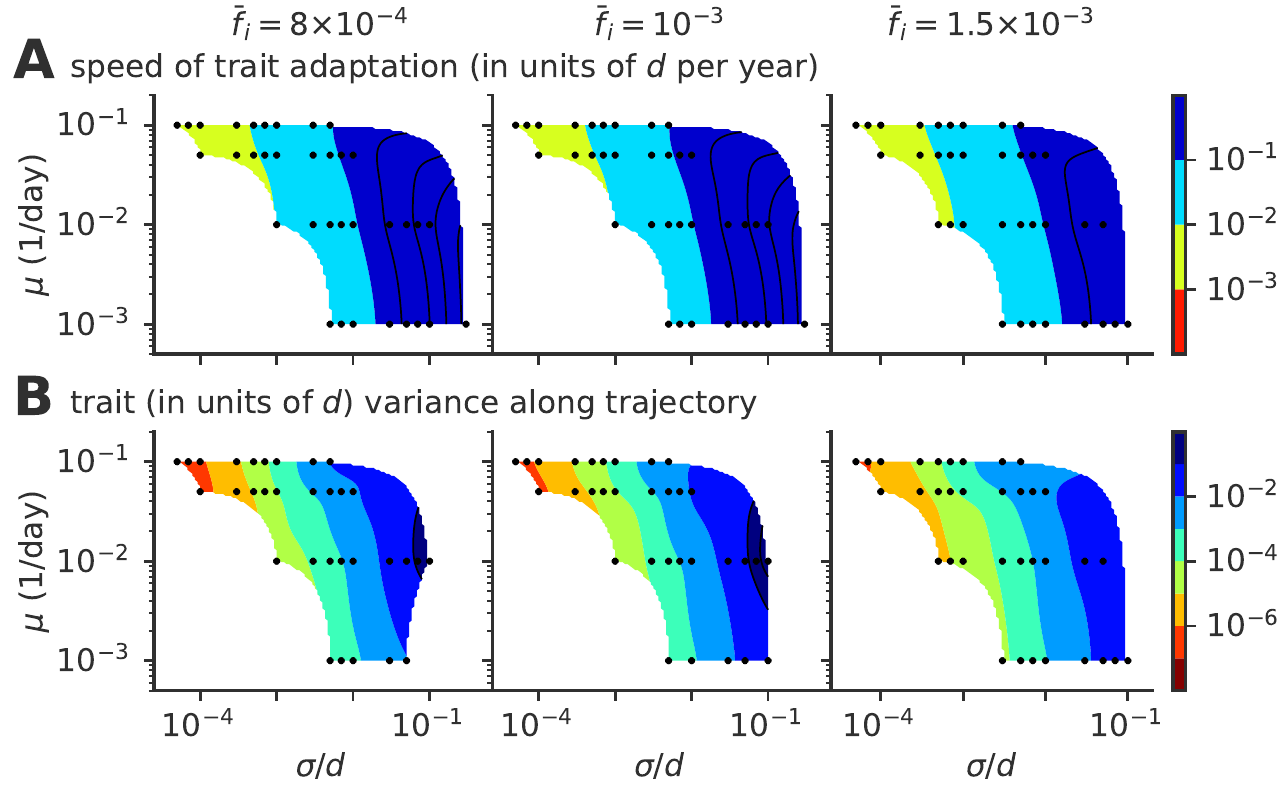}
\caption{\label{var_velFIX}{\bf Speed of adaptation and within-cluster diversity.} Same phase diagram as in Fig.~\ref{var_vel} of the main text but with constant fixed fraction of infected hosts $\bar f_i=8\cdot 10^{-4}$, $10^{-3}$, and $1.5\cdot 10^{-3}$ (from left to right). Phase diagrams  as a function of mutation rate $\mu$ and mutation jump rate $\sigma$ for (A) the average speed of the evolving viral clusters and (B) the phenotypic variance in the direction parallel to the direction of instantaneous mean adaptation.}
\end{figure*}

\begin{figure*}
\includegraphics[width=.7\linewidth]{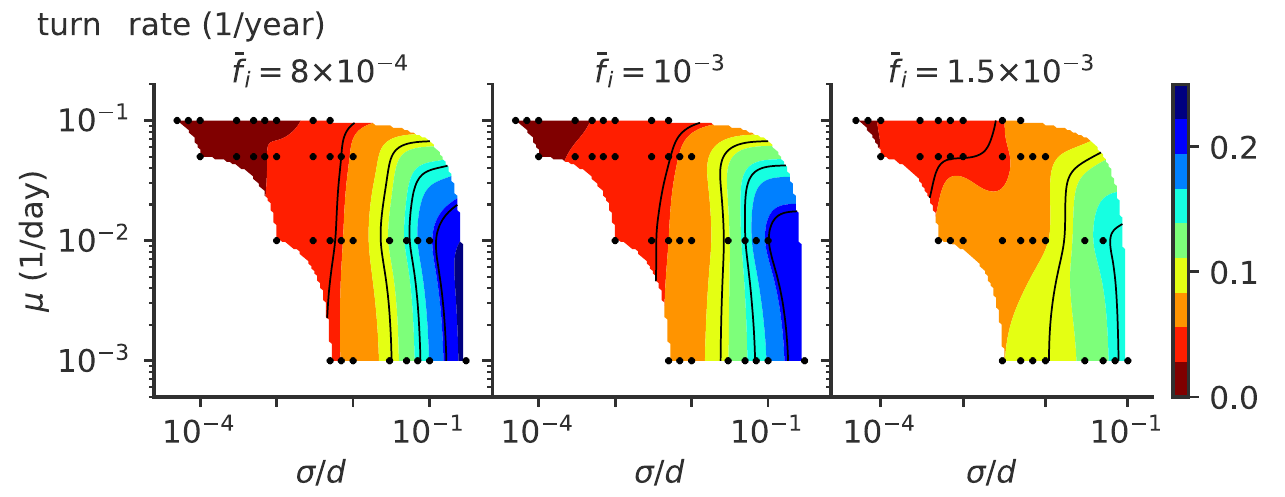}
\caption{\label{pers_fix}{\bf Persistence time.}  Same phase diagram as in Fig.~\ref{pers} of the main text but with constant fixed population size $\bar f_i=8\cdot 10^{-4}$, $10^{-3}$, and $1.5\cdot 10^{-3}$ (from left to right).
Phase diagrams as a function of mutation rate $\mu$ and mutation jump rate $\sigma$ for persistence time of the trajectories.}
\end{figure*}

\begin{figure*}
\includegraphics[width=.7\linewidth]{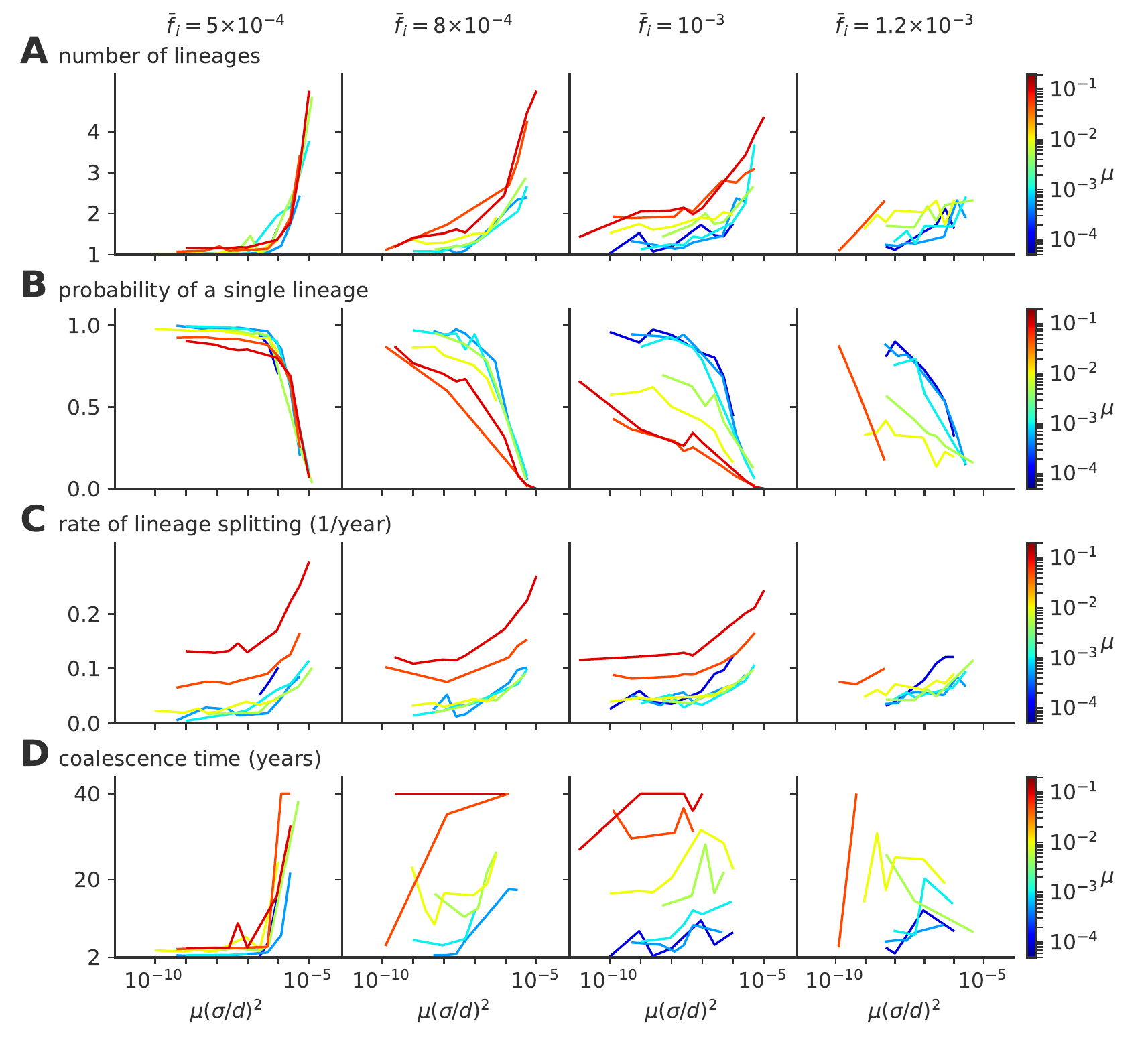}
\caption{\label{fig:scaling}{\bf Single- to multiple lineage transition a function of rescaled diffusivity $\mu\sigma^2$.} Same quantities as in Fig.~4 of the main text, but as a function of the effecive diffusivity $\mu\sigma^2$, showing absence of collapse as a function of that parameter for various values of the mutation rate $\mu$. (A) Average number of lineages, (B) fraction of evolution time where viruses are organized in a single lineage, (C) rate of lineages splitting (per lineage), and (D) average coalescence time.}
\end{figure*}

\end{document}